\documentclass[floatfix,aps,prl,twocolumn,groupedaddress,showpacs,superscriptaddress,preprintnumbers]{revtex4-1}

\usepackage[utf8]{inputenc}
\usepackage{graphicx}
\usepackage{slashed}
\usepackage{color}
\usepackage{amsmath}
\usepackage{amssymb}

\definecolor{mygreen}{rgb}{0,0.7,0}

\DeclareMathOperator{\tr}{\rm tr}

\def\eps{\epsilon}

\usepackage{multirow}
\usepackage{tabularx}
\newcolumntype{C}[1]{>{\hsize=#1\hsize\centering\arraybackslash}X}%

\newcolumntype{Z}{r<{\hspace{3mm}}}
\newcommand\mc[2]{\multicolumn{1}{>{\centering}p{#2}}{#1}} 

\begin{document}


\title{A first look at two-loop five-gluon scattering in QCD}


\author{Simon Badger}
\email[]{simon.d.badger@durham.ac.uk}
\affiliation{Institute for Particle Physics Phenomenology, Department of Physics, Durham University, Durham DH1 3LE, United Kingdom%
}

\author{Christian Br\o nnum-Hansen}
\affiliation{
Higgs Centre for Theoretical Physics, School of Physics and Astronomy, The University of Edinburgh, Edinburgh EH9 3JZ, Scotland, UK%
}

\author{Heribertus Bayu Hartanto}
\affiliation{Institute for Particle Physics Phenomenology, Department of Physics, Durham University, Durham DH1 3LE, United Kingdom%
}

\author{Tiziano Peraro}
\affiliation{PRISMA Cluster of Excellence, Johannes Gutenberg University, 55128 Mainz, Germany%
}


\date{\today}
\preprint{IPPP/17/95, Edinburgh 2017/27, MITP/17-094}

\begin{abstract}
We compute the leading colour contributions to five-gluon scattering at two loops in
massless QCD. The integrands of all independent helicity amplitudes are evaluated
using $d$-dimensional generalised unitarity cuts and finite field reconstruction techniques. Numerical
evaluation of the integral basis is performed with sector decomposition methods to obtain the first
benchmark results for all helicity configurations of a $2$ to $3$ scattering process in QCD.
\end{abstract}


\maketitle

\section{Introduction \label{sec:intro}}

As data continues to pour in from the LHC experiments, the precision of many
theoretical predictions for high energy scattering processes are being challenged
by experimental measurements.  While there has been remarkable progress in
Standard Model (SM) predictions for multi-particle final states at
next-to-leading-order (NLO) and $2\to2$ scattering processes at
next-to-next-to-leading order (NNLO), the computational complexity of $2\to3$
scattering processes at NNLO results in many important measurements being
currently (or in the near future) limited by theoretical uncertainties.
Pure gluon scattering at two loops in QCD is a key bottleneck in making such predictions
which have been known for $gg\to gg$ for more than 15 years~\cite{Glover:2001af,Bern:2002tk}.
The one-loop five-gluon amplitudes have been known since 1993~\cite{Bern:1993mq} and were among the first results from the on-shell methods that led
to the modern unitarity method~\cite{Bern:1994zx,Bern:1994cg}.

In this letter we demonstrate how new evaluation techniques based on generalised unitarity~\cite{Britto:2004nc,Bern:1997sc} and
integrand reduction~\cite{Ossola:2006us,Mastrolia:2011pr,Mastrolia:2012an,Mastrolia:2012wf,Mastrolia:2013kca,Badger:2012dp,Zhang:2012ce} can offer a solution to the traditional bottlenecks in these computations~
and present the first results for a complete set of planar five-gluon helicity amplitudes in QCD. The results extend previous results obtained for `all-plus' helicity amplitudes \cite{Badger:2013gxa,Badger:2015lda,Badger:2016ozq,Gehrmann:2015bfy,Dunbar:2016aux,Dunbar:2016cxp,Dunbar:2016gjb,Dunbar:2017nfy}. These on-shell techniques have also been explored in the context of maximal unitarity~\cite{Kosower:2011ty,CaronHuot:2012ab} and numerical unitarity~\cite{Abreu:2017xsl,Abreu:2017idw,Ita:2015tya} approaches to QCD amplitudes. Work in this area has received considerable interest due to the phenomenological importance of precision predictions for $2\to3$ scattering. Efforts to complete the unknown two-loop amplitudes for processes such as $pp\to3$ jets, $pp\to H+2$ jets or $pp\to \gamma\gamma+$jet have been further motivated by the recent analytic computations of the planar master integrals (MIs)~\cite{Papadopoulos:2015jft,Gehrmann:2015bfy} using new differential equation techniques~\cite{Henn:2013pwa,Papadopoulos:2014lla}.

Our approach exploits a parametrisation of the multi-particle kinematics with rational functions combined with
numerical evaluation over finite fields~\cite{Peraro:2016wsq} to avoid the large intermediate algebraic expressions that traditionally appear.
The rational parametrisation of the external kinematics is provided by momentum twistor
coordinates~\cite{Hodges:2009hk}.

\section{Integrand parametrisation and reconstruction \label{sec:integrand}}

We define the unrenormalised leading-colour (planar) five-gluon amplitudes using the simple trace basis:
\begin{align}
  \mathcal{A}^{(L)}&(1,2,3,4,5) = n^L g_s^3 \sum_{\sigma \in S_5/Z_5} \tr \left(
  T^{a_{\sigma(1)}}\cdots T^{a_{\sigma(5)}} \right) \nonumber\\& \times
  A^{(L)}\left(\sigma(1),\sigma(2),\sigma(3),\sigma(4),\sigma(5) \right),
  \label{eq:lcampdef}
\end{align}
where $T^a$ are the fundamental generators of $SU(N_c)$ and $S_5/Z_5$ are all
noncyclic permutations of the external particles.  The overall normalisation is $n
= m_\eps N_c \alpha_s/(4\pi)$ where $\alpha_s = g_s^2/(4\pi)$ is the strong
coupling constant and $m_\eps=i (4\pi)^{\eps} e^{-\eps\gamma_E}$ ($\gamma_E$ is
the Euler–Mascheroni constant).  The $L$-loop partial amplitude
$A^{(L)}$ can be constructed from colour ordered Feynman diagrams. In this article we will compute
the pure gluonic contributions to these amplitudes at two loops including the dependence on the
spin dimension, $d_s$. Results in the ’t Hooft-Veltman (tHV) and four-dimensional-helicity (FDH)
schemes can be obtained by setting $d_s=4-2\eps$ and $d_s=4$ respectively~\cite{Bern:2002zk}.

\begin{figure}[t]
  \includegraphics[width=0.3\textwidth]{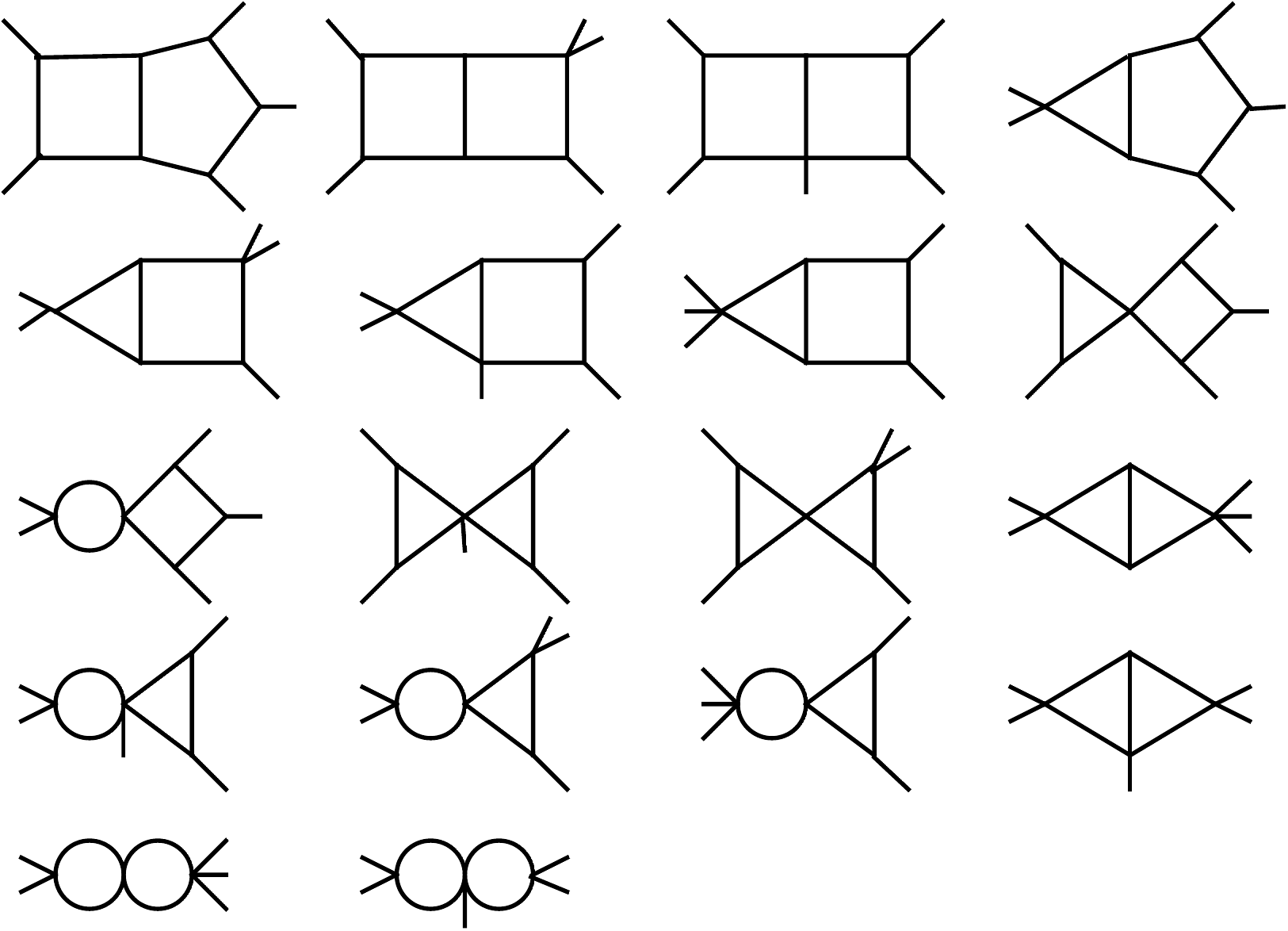}%
  \caption{The 18 distinct topologies extractable from (1-loop)${}^2$ cuts. \label{fig:topo_1Lsq}}
\end{figure}

\begin{figure}[b]
  \includegraphics[width=0.35\textwidth]{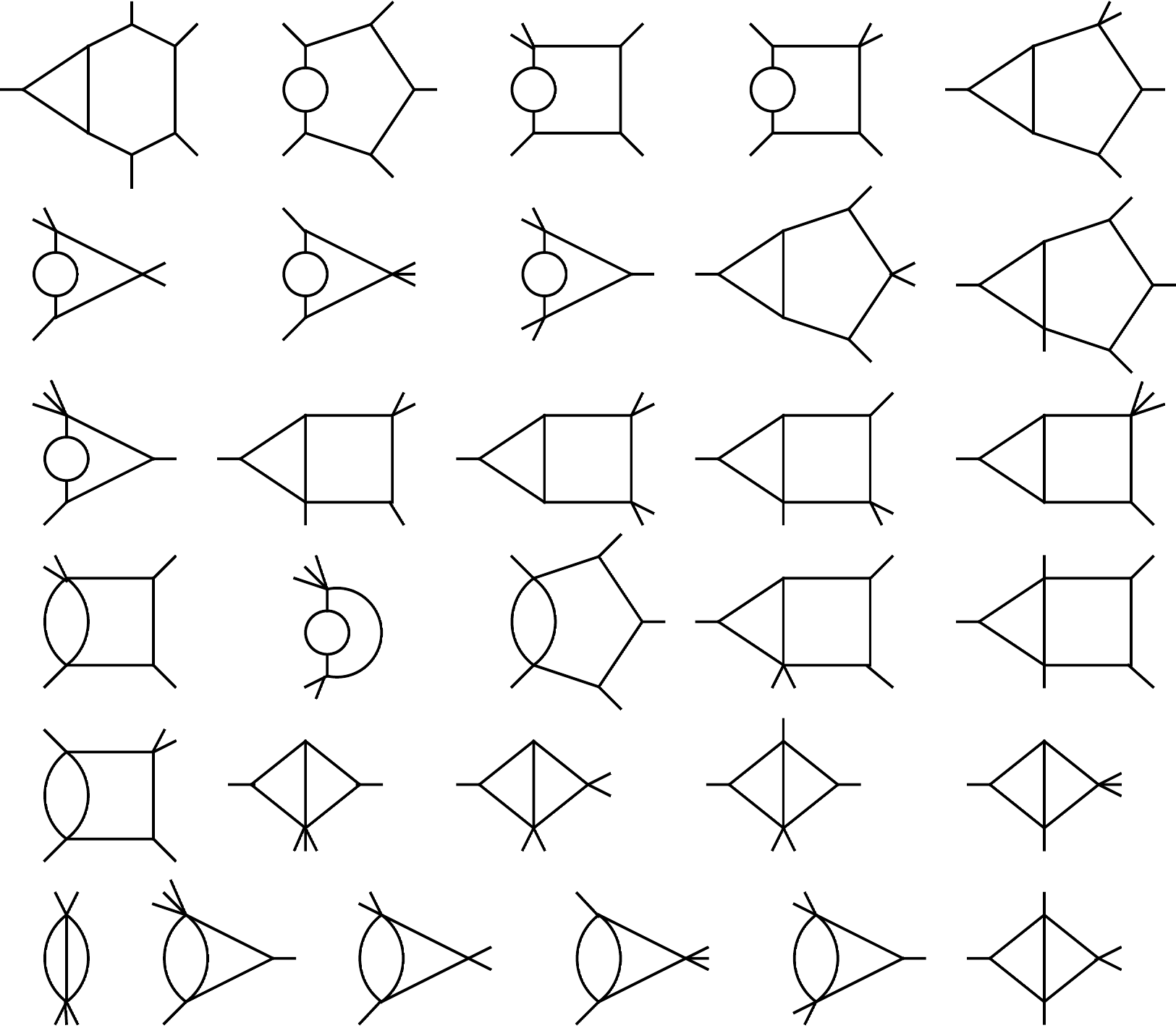}%
  \caption{31 distinct topologies extractable from 2-loop cuts. \label{fig:topo_2L}}
\end{figure}

\begin{figure}[b]
  \includegraphics[width=0.25\textwidth]{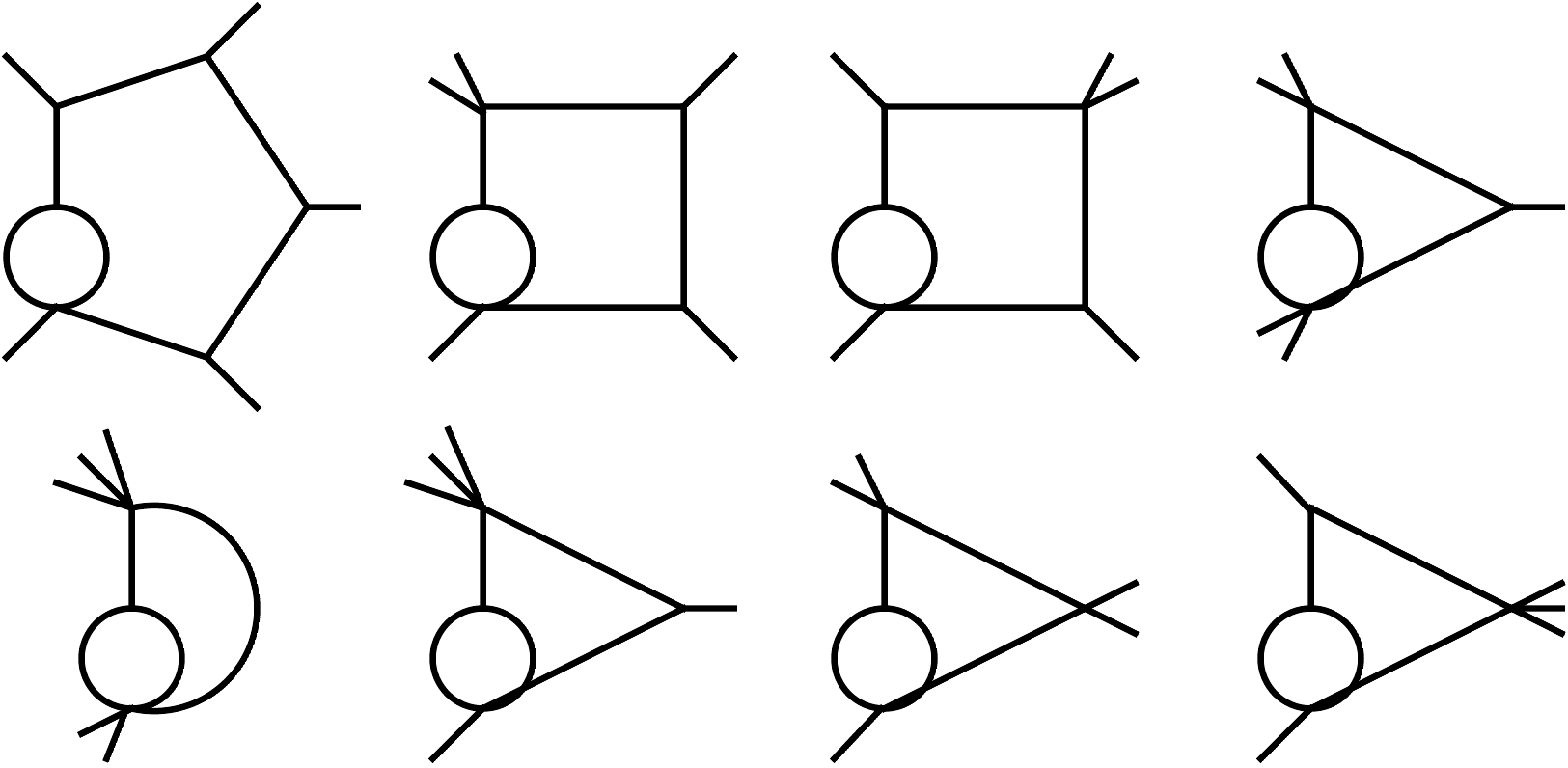}%
  \caption{The 8 distinct topologies with divergent cuts that must be computed simultaneously with subtopologies \label{fig:topo_2Ldiv}}
\end{figure}

The integrand of the ordered partial amplitudes can be parametrised in terms of irreducible
numerators, $\Delta$,
\begin{equation}
  A^{(2)}\left(1,2,3,4,5\right) = \int
  [dk_1][dk_2]
  \, \sum_T
  \frac{\Delta_{T}(\{k\},\{p\})}{\prod_{\alpha \in T} D_\alpha},
  \label{eq:integrand}
\end{equation}
where $\{k\}=\{k_1,k_2\}$ are the $(d=4-2\eps)$-dimensional loop momenta, $T$
are the set of independent topologies and $\{p\} = \{1,2,3,4,5\}$ are the ordered
external momenta. The measure is $[dk_i] = -i \pi^{-d/2} e^{\epsilon\gamma_E} d^{4-2\eps} k_i$ and the
index $\alpha$ runs over the set of propagators associated with the topology
$T$. Our planar five-gluon amplitudes are built from 425 irreducible numerators
with 57 distinct topologies. 18 of these 57 can be extracted from the
(1-loop)${}^2$ cut configurations as shown in Fig.~\ref{fig:topo_1Lsq}.
This means that all topologies with an additional propagator including $k_1+k_2$
are computed simultaneously with the (1-loop)${}^2$ cuts.  This is more efficient since
the parametrisations of the cut loop momentum solutions are much simpler. The remaining 39 can be
extracted from a further 31 configurations shown in Fig.~\ref{fig:topo_2L}. The 8 topologies shown in Fig.~\ref{fig:topo_2Ldiv} have
divergent maximal cuts and are extracted simultaneously with sub-topologies
within the set of 31 2-loop cuts.

The construction of an integrand basis has been discussed before using the
language of computational algebraic geometry through polynomial division over a
Gr\"obner basis~\cite{Zhang:2012ce,Mastrolia:2012an}. In this work we took a simpler approach which did not rely on
the computation of a Gr\"obner basis, instead relying on the inversion of a
linear system which can be performed efficiently with finite field
reconstruction methods. We begin by expanding the loop momenta around a basis
of external momenta and transverse directions (similarly to the methods of Van Neerven and Vermaseren~\cite{vanNeerven:1983vr}),
\begin{equation}
  k_i^\mu = k_{\parallel,i}^\mu + k_{\perp,i}^\mu,
  \label{eq:transdecomp}
\end{equation}
where $k_{\parallel}$ lives in the physical space spanned by the
external momenta of the topology and $k_{\perp}$
lives in the
transverse space.  We further decompose the transverse space into four dimensional and
$(-2\eps)$ dimensional spaces,
$k_{\perp, i} = k^{[4]}_{\perp, i}+k^{[-2\eps]}_{\perp, i}$.  The
size of the $4$-d transverse space (which we will call the spurious space)
has dimension $d_{\perp,[4]} = 4- d_{\parallel}$ where $d_{\parallel}$ is equal to the number of independent momenta
entering the vertices of the topology, up to a maximum value of four.
We choose a spanning basis $v$ for the physical space of each
topology $k_{\parallel,i}^\mu = \sum_{j=1}^{d_{\parallel}} a_{ij} v_j^\mu$ and
a basis $w$ for the spurious space $k_{\perp,i}^{\mu,[4]} =
\sum_{j=1}^{d_{\perp,[4]}} b_{ij} w_j^\mu$, with $v_i.w_j=0$.

The coefficients in the physical space $k_\parallel$ are functions of
the $a_{ij}(k_i) \equiv a_{ij}(\{D\},\{k.q\})$ where $D$ are the inverse propagators
and $k_i.q_j$ are the physical space irreducible scalar products (ISPs) for a given topology, where $q_j$ are suitable linear combinations of external momenta. The
coefficients in the spurious and $(-2\eps)$-d spaces are functions of
additional ISPs $k_i.w_j$ and $\mu_{ij}=-k^{[-2\eps]}_{\perp, i}.k^{[-2\eps]}_{\perp, j}$.  Having completed this decomposition
we find relations between monomials in the ISPs by expanding Eq.
\eqref{eq:transdecomp},
\begin{equation}
  \mu_{ij} = k_i.k_j - k_{\parallel,i}.k_{\parallel,j}  - k^{[4]}_{\perp,i}.k^{[4]}_{\perp,j}.\label{eq:removemu}
\end{equation}
From this equation it is easy to obtain a valid basis of monomials for each
irreducible numerator of a dimensionally regulated amplitude by using Eq.
\eqref{eq:removemu} to remove dependence on the extra dimensional ISPs. This
basis is just the most general polynomial in the ISPs $k_i.q_j$ and $k_i.w_j$
where the power counting is restricted by the renormalizability
constraints~\footnote{The renormalizability constraints restrict the maximum
rank in the loop momenta that can appear in $\Delta$. This is discussed in more
detail in the literature on multi-loop integrand reduction}.

This basis is trivial to obtain without polynomial division but results in high rank tensor
integrals with a complicated infrared (IR) pole structure. Instead we prefer to map
to a new basis which prefers to keep monomials in $\mu_{ij}$ in the numerator and
make the $\eps\to0$ limit easier to perform. The map to the new basis is performed in four steps:
1) write down a complete set of monomials in $k_i.q_j$, $k_i.w_j$ and
$\mu_{ij}$ obeying the power counting restrictions. 2) Order the monomials with
respect to a set of reasonable criteria (for example prefer lower rank monomials
or prefer monomials proportional to $\mu_{ij}$). 3) Map all monomials onto the
simple basis and construct a linear system according to the ordering of
variables. 4) Solve the linear system for the independent monomials in the new
basis. The result of this procedure is a process independent basis of monomials whose coefficients
can be fixed from unitarity cuts in six dimensions. We take a top-down, OPP-like, approach to solving the
complete system using information from previously computed cuts to remove known poles from
the factorised product of tree amplitudes using the six-dimensional spinor-helicity formalism~\cite{Cheung:2009dc}.
The product of tree amplitudes is efficiently evaluated by sewing together Berends-Giele currents~\cite{Berends:1987me} as described in Ref.~\cite{Peraro:2016wsq}.

After completing the integrand level reconstruction, the remaining transverse integration
must be performed to obtain a form compatible with traditional
integration-by-parts (IBP) relations. Following a recent approach~\cite{Mastrolia:2016dhn}, we have two options in order to achieve this: 1) to integrate
the full transverse space to remove $k_i.w_j$ and $\mu_{ij}$ introducing
dependence in $\eps$ into the integral coefficients or, 2) integrate only over
the spurious space retaining $\mu_{ij}$ dependence which can subsequently be
removed through dimension shifting identities.
In this work we have taken the second approach
since it turned out to have better numerical stability to use dimension shifted integrals
instead of high rank tensor integrals.

In either case the tensor structure in the transverse space can only involve the metric
tensor $g^{\mu\nu}_{\perp}$ (or $g^{\mu\nu}_{\perp,[4]},g^{\mu\nu}_{\perp,[-2\eps]}$
depending on the particular transverse space being integrated out). This makes the tensor decomposition
for non-vanishing integrals in the spurious space rather simple. Further examples of this technique
can be found in Ref.~\cite{Mastrolia:2016dhn}.

We build integration identities and certain symmetry relations (for
example $k_1 \leftrightarrow k_2$ in the 3-propagator sunrise topology) into the integrand basis by
using them to create spurious numerators. For example, rather than
fitting the coefficient of $(k_1.w_2)^2$ we replace it with the
function
\begin{equation}
  (k_1.w_2)^2 \longrightarrow (k_1.w_2)^2 - \tfrac{w_2^2}{d_{\perp,[4]}} k^{[4]}_{\perp,1}.k^{[4]}_{\perp,1},
\end{equation}
which will integrate to zero. In Tab.~\ref{tab:coeffcount}
we summarise the result of our fit to unitarity cuts listing the number of non-zero
coefficients at the integrand level before and after performing the integration over the spurious space.
Cuts with scalar loops are required for the reduction from $6$ to $4-2\eps$ dimensions. We perform the fit taking into account the individual contribution of these scalar loops in order to reconstruct the
dependence of the numerator on the spin dimension $d_s$. Setting $d_s=2$ gives a supersymmetric limit in which the highest rank tensor integrals do not appear in the amplitudes. We use a polynomial expansion of the
integrand in $(d_s-2)$ to separate the coefficients into terms of increasing complexity.
The fit can be performed efficiently using rational numerics for each phase
space point and in most cases it was possible to obtain completely analytic
expressions for the integrands of the helicity amplitudes using modest computing resources.

\renewcommand{\arraystretch}{1.2}
\begin{table}[b]
\begin{tabular}{ccZZZ}
  \hline
  helicity & flavour & \mc{non-zero coefficients}{1.6cm} & \mc{non-spurious coefficients}{1.9cm} & \mc{contributions @ $\mathcal{O}(\eps^0)$}{2cm} \\
  \hline
  \multirow{3}{*}{${}_{+++++}$} & $(d_s-2)^0$ & 50    & 50   & 0\\
                                & $(d_s-2)^1$ & 175   & 165  & 50 \\
                                & $(d_s-2)^2$ & 320   & 90   & 60 \\
  \hline
  \multirow{3}{*}{${}_{-++++}$} & $(d_s-2)^0$ & 1153  & 761  & 405 \\
                                & $(d_s-2)^1$ & 8745  & 4020 & 3436 \\
                                & $(d_s-2)^2$ & 1037  & 100  & 68 \\
  \hline
  \multirow{3}{*}{${}_{--+++}$} & $(d_s-2)^0$ & 2234  & 1267 & 976 \\
                                & $(d_s-2)^1$ & 11844 & 5342 & 4659 \\
                                & $(d_s-2)^2$ & 1641  & 71   & 48 \\
  \hline
  \multirow{3}{*}{${}_{-+-++}$} & $(d_s-2)^0$ & 3137  & 1732 & 1335  \\
                                & $(d_s-2)^1$ & 15282 & 6654 & 5734 \\
                                & $(d_s-2)^2$ & 3639  & 47   & 32 \\
  \hline
\end{tabular}
  \caption{\label{tab:coeffcount} The number of non-zero coefficients found at
  the integrand level both before (`non-zero') and after (`non-spurious')
  removing monomials which integrate to zero. Last column (`contributions @ $\mathcal{O}(\eps^0)$') gives the number of coefficients contributing to the finite part. Each helicity amplitude is split into
  the components of $d_s-2$.}
\end{table}

\section{Numerical evaluation \label{sec:numerics}}

The unitarity based method outlined above has been complemented by an approach based on numerical evaluation of
Feynman diagrams to determine the coefficients of independent monomial bases.
Both of these methods use a momentum twistor~\cite{Hodges:2009hk} parametrisation of the external
kinematics to obtain a rational numerical phase-space point. This is extremely important since in order
to make use of the finite field reconstruction methods our numerical algorithm must be free of all square roots~\cite{Wang:1981:PAU:800206.806398,Wang:1982:PRR:1089292.1089293,Trager:2006:1145768,RISC3778}.
The parametrisation in this case was chosen (somewhat arbitrarily) to be,
\begin{align}
  Z = \begin{pmatrix}
    1 & 0 & \tfrac{1}{x_1} & \tfrac{1+x_2}{x_1x_2} & \tfrac{1+x_3(1+x_2)}{x_1x_2x_3} \\
    0 & 1 & 1 & 1 & 1 \\
    0 & 0 & 0 & \tfrac{x_4}{x_2} & 1 \\
    0 & 0 & 1 & 1 & \tfrac{x_4-x_5}{x_4}
  \end{pmatrix},
  \label{eq:mtwistorparam}
\end{align}
where the columns give the 4-component momentum twistors of the 5 external particles (see, for example, Appendix A of Ref.~\cite{Badger:2013gxa} for more details). These methods
have been implemented using a combination of tools including \textsc{Qgraf}~\cite{Nogueira:1991ex}, \textsc{Form}~\cite{Kuipers:2012rf,Ruijl:2017dtg}, \textsc{Mathematica} and a private
implementation of the finite field reconstruction method~\cite{Peraro:2016wsq}.

We have validated our setup on a number of known cases. Firstly, we have
reproduced integrand level expressions for the `all-plus' helicity sector~\cite{Badger:2013gxa} and
against the known integrands in $\mathcal{N}=4$ Super-Yang-Mills theory~\cite{Bern:2006vw}. The
latter check was obtained by computing all fermion and (complex)-scalar loop
contributions and subsequently setting $n_f = \mathcal{N}$ and $n_s = \mathcal{N}-1$. We
also have performed gauge invariance checks at the integrand level using the Feynman diagram setup.

To obtain a numerical value for the complete amplitude after integration we
perform a sector decomposition of the basis integrals combined with Monte Carlo integration. After applying dimension shifting
relations~\cite{Tarasov:1996br,Bern:2002tk,Lee:2009dh} to rewrite the extra-dimensional ISPs as standard integrals we
processed the full set of integrals using both \textsc{Fiesta}~\cite{Smirnov:2015mct} and \textsc{pySecDec}~\cite{Borowka:2017idc} packages.
This setup was validated with the four-gluon helicity amplitudes and
cross-checked against results in the literature~\cite{Abreu:2017xsl}.
Simple topologies with $2\to2$ kinematics were reduced to the known MIs of Ref.~\cite{Gehrmann:2000zt} using IBPs from \textsc{Fire5}~\cite{Smirnov:2014hma} and
\textsc{Reduze2}~\cite{vonManteuffel:2012np} and dimensional recurrence relations from \textsc{LiteRed}~\cite{Lee:2012cn}. This gave a substantial improvement in the numerical accuracy.

\renewcommand{\arraystretch}{1.3}
\begin{table}[t]
  \begin{tabularx}{0.48\textwidth}{|C{1.1}|C{0.5}|C{0.9}|C{0.8}|C{1}|C{1.7}|}
    \hline
    &  $\eps^{-4}$ & $\eps^{-3}$ & $\eps^{-2}$ & $\eps^{-1}$ & $\eps^{0}$ \\
    \hline
    $\widehat{A}^{(2),[0]}_{--+++}$ & 12.5 & 27.7526 & -23.773 & -168.117 & -175.207$\pm$0.004 \\
              $P^{(2),[0]}_{--+++}$ & 12.5 & 27.7526 & -23.773 & -168.116 & \quad--- \\
    \hline
    $\widehat{A}^{(2),[0]}_{-+-++}$ & 12.5 & 27.7526 & 2.5029 & -35.8094 & 69.661$\pm$0.009 \\
              $P^{(2),[0]}_{-+-++}$ & 12.5 & 27.7526 & 2.5028 & -35.8086 & \quad--- \\
    \hline
  \end{tabularx}
  \caption{\label{tab:numericalresults1} The numerical evaluation of $\widehat{A}^{(2),[0]}(1,2,3,4,5)$ using $\{x_1 = -1, x_2 = 79/90,
x_3 = 16/61, x_4 = 37/78, x_5 = 83/102\}$ in Eq.\eqref{eq:mtwistorparam}. The comparison with the universal pole structure, $P$, is shown.
  The \texttt{+++++} and \texttt{-++++} amplitudes vanish to $\mathcal{O}(\eps)$ for this $(d_s-2)^0$ component.}
\end{table}

\begin{table}[b]
  \begin{tabularx}{0.48\textwidth}{|C{1.1}|C{0.5}|C{0.8}|C{0.9}|C{1}|C{1.7}|}
    \hline
    &  $\eps^{-4}$ & $\eps^{-3}$ & $\eps^{-2}$ & $\eps^{-1}$ & $\eps^{0}$ \\
    \hline
    $\widehat{A}^{(2),[1]}_{+++++}$ & 0 & 0.0000 & -2.5000 & -6.4324 & -5.311$\pm$0.000 \\
              $P^{(2),[1]}_{+++++}$ & 0 & 0 & -2.5000 & -6.4324 & \quad--- \\
    \hline
    $\widehat{A}^{(2),[1]}_{-++++}$ & 0 & 0.0000 & -2.5000 & -12.749 & -22.098$\pm$0.030 \\
              $P^{(2),[1]}_{-++++}$ & 0 & 0 & -2.5000 & -12.749 & \quad--- \\
    \hline
    $\widehat{A}^{(2),[1]}_{--+++}$ & 0 & -0.6250 & -1.8175 & -0.4871 & 3.127$\pm$0.030 \\
              $P^{(2),[1]}_{--+++}$ & 0 & -0.6250 & -1.8175 & -0.4869 & \quad--- \\
    \hline
    $\widehat{A}^{(2),[1]}_{-+-++}$ & 0 & -0.6249 & -2.7761 & -5.0017 & 0.172$\pm$0.030 \\
              $P^{(2),[1]}_{-+-++}$ & 0 & -0.6250 & -2.7759 & -5.0018 & \quad--- \\
    \hline
  \end{tabularx}
  \caption{\label{tab:numericalresults2} The numerical evaluation of $\widehat{A}^{(2),[1]}(1,2,3,4,5)$ and comparison with the universal pole structure, $P$, at the same
  kinematic point of Tab. \ref{tab:numericalresults1}.}
\end{table}

The results for evaluation at a specific phase-space point are given in Tables
\ref{tab:numericalresults1} and \ref{tab:numericalresults2} for the amplitudes
\begin{equation}
\widehat{A}^{(2),[i]}_{\lambda_1\lambda_2\lambda_3\lambda_4\lambda_5} =%
\frac{A^{(2),[i]}(1^{\lambda_1},2^{\lambda_2},3^{\lambda_3},4^{\lambda_4},5^{\lambda_5})}{%
  A^{\rm LO}(1^{\lambda_1},2^{\lambda_2},3^{\lambda_3},4^{\lambda_4},5^{\lambda_5})},
\end{equation}
with helicities $\lambda_i$ and $A^{(2)} = \sum_{i=0}^2(d_s-2)^i A^{(2),[i]}$.
The leading order amplitudes $A^{\rm LO}$ are the tree-level for the \texttt{--+++}
and \texttt{-+-++} and rational one-loop amplitudes for the \texttt{+++++} and \texttt{-++++}. The
finite (1-loop)${}^2$ configuration $A^{(2),[2]}$ is presented in Tab.
\ref{tab:numericalresults3}. Numerical accuracy is not an issue here since the
integrand level reduction already leads to a basis of one-loop MIs. In addition
we find complete agreement with the finite part of the known integrated `all-plus' amplitude \cite{Gehrmann:2015bfy}.

In cases where the $\eps$ pole structure of the amplitudes is non-trivial we
compared with the known universal IR
structure~\cite{Catani:1998bh,Becher:2009qa,Becher:2009cu,Gardi:2009qi}
including the dependence on $d_s$ extracted from the FDH scheme results
\cite{Gnendiger:2014nxa}. The leading pole in $1/\eps^4$ was verified
analytically and is therefore quoted exactly in Tabs.~\ref{tab:numericalresults1} and \ref{tab:numericalresults2}.  By comparing the
agreement in the poles between the $(d_s-2)^0$ and $(d_s-2)^1$ we clearly
see the effect of the highest rank tensor integrals which only appear in the
latter case. We find convincing agreement between the poles and our amplitudes
within the numerical integration error~\footnote{The uncertainty on the finite
terms in Tabs. \ref{tab:numericalresults1} and \ref{tab:numericalresults2} is a
rough estimate made by comparing \textsc{Fiesta} evaluations with different numbers of
sample points.}. Since the full amplitude is the sum of all three parts we see in this case that the simple
$(d_s-2)^0$ part dominates and the complete amplitude is evaluated with sub-percent level accuracy. This feature is probably not generic for the whole
phase-space however.

\begin{table}[t]
  \begin{tabularx}{0.45\textwidth}{|C{0.08}|C{0.23}|C{0.23}|C{0.23}|C{0.23}|}
    \hline
    & $\widehat{A}^{(2),[2]}_{+++++}$ & $\widehat{A}^{(2),[2]}_{-++++}$ & $\widehat{A}^{(2),[2]}_{--+++}$ & $\widehat{A}^{(2),[2]}_{-+-++}$ \\
    \hline
    $\eps^0$ & 3.6255 & -0.0664 & 0.2056 & 0.0269 \\
    \hline
  \end{tabularx}
  \caption{\label{tab:numericalresults3} The numerical evaluation of finite $\widehat{A}^{(2),[2]}(1,2,3,4,5)$ helicity amplitudes at the same
  kinematic point of Tab. \ref{tab:numericalresults1}. As only one-loop integrals are required for these amplitudes the integration error is negligible.}
\end{table}

\section{Conclusions \label{sec:conclusions}}

The techniques presented in this letter have allowed the first look at a set of five-point
two-loop helicity amplitudes with phenomenological relevance for LHC experiments. We have found
that unitarity cutting methods in six dimensions can be combined with finite field reconstruction
techniques to compute multi-scale dimensionally regulated two-loop amplitudes in QCD. In many cases
it was possible to obtain completely analytic expressions for the
integrands of the helicity amplitudes.

While a lot of effort was taken to find manageable expressions, the final integrand form
was still extremely large and significantly more challenging than the previously known
`all-plus' helicity configuration. One obvious next step is to include a full set of integration-by-parts identities and reduce the amplitude onto a basis of
analytically computed MIs. Promising new approaches that use finite field reconstruction~\cite{vonManteuffel:2014ixa}
or algebraic geometry analyses~\cite{Gluza:2010ws,Ita:2015tya,Larsen:2015ped,Georgoudis:2016wff,Bern:2017gdk} could make this possible in the near future.
We expect there will be other ways to improve the integrand form by using canonical bases~\cite{Henn:2013pwa} and local integrand representations~\cite{ArkaniHamed:2010kv,ArkaniHamed:2010gh,Bourjaily:2017wjl}
though at the present time more work is needed to investigate these approaches.

While there clearly remains a long list of tasks to be completed before predictions of $2\to 3$ scattering
at NNLO in QCD become a reality, the work presented here is the first example of the evaluation of one of the key
ingredients. We hope that the techniques and benchmark results presented here will provide a platform towards this final goal.

\begin{acknowledgments}
  We thank Fabrizio Caola, Johannes Henn, Claude Duhr, Donal O'Connell, Nigel Glover, Adriano Lo Presti, Francesco Buciuni, and Johannes Schlenk
  for many stimulating discussions. We would particularly like to thank Claude Duhr for providing computer readable expressions for the integrals in Ref.~\cite{Gehrmann:2000zt}. It is also a pleasure to thank the organisers of the AMPDEV2017 program at the MITP, Mainz and also the organisers of the
  ``Automated, Resummed and Effective: Precision Computations for the LHC and Beyond" and ``Mathematics and Physics of Scattering Amplitudes" programs at the MIAPP centre, Munich
  for creating stimulating working environments while this work was on-going. This work is supported by the STFC fellowship ST/L004925/1 and grant ST/M004104/1.   The work of T.P.\ has received funding from the European Research Council (ERC) under the European Union’s Horizon 2020 research and innovation programme (grant agreement No 725110). We are also indebted to Samuel Abreu, Fernando Febres Cordero, Harald Ita, Ben Page and Mao Zeng for pointing out typographical errors in a previous version of this letter.

\end{acknowledgments}

\bibliography{2L5gPlanar}

\end{document}